\documentclass[11pt]{article}

\usepackage{graphicx}%
\usepackage{svg} 
\usepackage{multirow}%
\usepackage{amsmath,amssymb,amsfonts}%
\usepackage{amsthm}%
\usepackage{mathrsfs}%
\usepackage[title]{appendix}%
\usepackage{xcolor}%
\usepackage{textcomp}%
\usepackage{manyfoot}%
\usepackage{booktabs}%
\usepackage{algorithm}%
\usepackage{algorithmicx}%
\usepackage{algpseudocode}%
\usepackage{listings}%
\usepackage[T1]{fontenc}
\usepackage{orcidlink}
\usepackage{authblk}      
\usepackage{hyperref}    

\newcommand{\bmhead}[1]{\subsubsection*{#1}}

\author[1]{Piotr Bajger \orcidlink{0000-0002-9487-8476}}
\author[1]{Roman Dusek \orcidlink{0009-0009-6388-9488}}
\author[1]{Krzysztof Galias \orcidlink{0009-0009-6388-9488}}
\author[1]{Paweł Młyniec}
\author[1,2]{Aleksander Wawer \orcidlink{0000-0002-7081-9797}}
\author[1,3]{Paweł Zawistowski \orcidlink{0000-0002-0273-7060}}

\affil[1]{Allegro sp. z o.o., ul. Wierzbięcice 1B, Poznań, 61-659, Poland}
\affil[2]{Institute of Computer Science, Polish Academy of Sciences, Jana Kazimierza 5, Warsaw, 01-248 Poland}
\affil[3]{Warsaw University of Technology, Pl. Politechniki 1, Warsaw, 00-661 Poland}

\title{MLPlatt: Simple Calibration Framework for Ranking Models}

\begin{document}

\maketitle


\begin{abstract}
Ranking models are extensively used in e-commerce for relevance estimation. These models often suffer from poor interpretability and no scale calibration, particularly when trained with typical ranking loss functions. This paper addresses the problem of post-hoc calibration of ranking models.
We introduce MLPlatt: a simple yet effective ranking model calibration method that preserves the item ordering and converts ranker outputs to interpretable click-through rate (CTR) probabilities usable in downstream tasks.
The method is context-aware by design and achieves good calibration metrics globally, and within strata corresponding to different values of a selected categorical field (such as user country or device), which is often important from a business perspective of an E-commerce platform.
We demonstrate the superiority of MLPlatt over existing approaches on two datasets, achieving an improvement of over 10\% in F-ECE (Field Expected Calibration Error) compared to other methods.
Most importantly, we show that high-quality calibration can be achieved without compromising the ranking quality.
\end{abstract}




\maketitle

\section{Introduction}\label{sec:intro}

Machine learning (ML) is often applied in scenarios where the predictions generated are, at least to some extent, uncertain. In a binary classification model, for example, one may be concerned not only with correctly predicting the class, but also with correctly predicting the probability that an item belongs to that class. A model whose output can be interpreted as a probability is said to be calibrated.

Research shows that many popular ML architectures (especially deep learning methods) often produce uncalibrated outputs \cite{Korb1999CalibrationAT,Guo2017OnCO,ovadia2019trust}. One remedy to this problem is to utilise calibration techniques as a corrective measure. These methods bring the raw model predictions closer to the underlying probability distribution. Applying such techniques for ranking models poses some unique challenges, such as preserving the ordering of items in a ranked list so as not to compromise the ranking quality.

\subsection{Ranking and CTR models}

Ranking models and Click-Through-Rate (CTR) prediction models are fundamental components of modern information retrieval (IR) systems. They influence how search engines, recommendation systems, and online marketplaces order and present content. Ranking models aim to estimate the relevance of items, such as documents, web pages, or products. Given a user query and context, ranking models output an ordered list of items that maximizes user satisfaction and utility. CTR prediction models, by contrast, estimate the probability of a user clicking on a given item. These two types of models frequently operate together: ranking models define the initial relevance ordering, while CTR models refine it using behavioral signals from historical user interactions.

To solve CTR prediction and ranking problems models with different architectures and loss functions are typically used. Both goals are treated as independent of each other. As observed in~\cite{ULTR_2024}, gains in click prediction do not necessarily translate to enhanced ranking performance. Conversely, ranking models perform poorly on regression metrics. As a result, score-sensitive CTR models lean toward regression-only approaches even if they are suboptimal for user-facing ranking metrics~\cite{Bai2022RegressionCL}. Most importantly, ranking models typically do not output probabilities. Commonly used ranking list-wise or pair-wise loss functions are invariant to rank-preserving input transformations \cite{Bai2022RegressionCL}. In other words, ranker scores are only interpretable in relation to other scores within the same listing. This makes the raw ranking model output not only unusable for downstream tasks but also difficult to interpret and explain.

In this study, we aim to convert ranking model scores into CTR probabilities in a monotonic manner, making the ranking scores interpretable while preserving item ordering.

\subsection{Motivation}

Typical real-world IR pipelines consist of a search engine  followed by a ranker. They focus on estimating relevance based on textual or contextual features. CTR prediction models play a complementary role: they capture behavioral intent by predicting the likelihood that a user will engage with a given result.

CTR models are often used for ranking refinement, where predicted click probabilities help reorder candidate results to maximize user engagement. In online advertising, CTR models estimate the expected value of impressions, directly influencing ad selection and bidding strategies.

When ranker outputs are calibrated to reflect true click probabilities, a single model can jointly serve both ranking, engagement prediction, and advertising scoring, simplifying the overall system architecture. This probabilistic interpretation enhances transparency and consistency: the same score that determines the item ordering also estimates the probability of user interaction. In retrieval pipelines, such unified models may reduce the need for separate CTR estimators, lowering latency and system complexity. 

In ad ranking, a calibrated ranker can balance relevance and expected revenue, since predicted CTRs can be combined with bid values. Moreover, probabilistic outputs improve interpretability, enable probabilistic fusion of signals across stages (e.g., retrieval and reranking), and support unified evaluation metrics such as log-loss and calibration error alongside ranking metrics like NDCG.

 One notable example of such a fusion of signals, which specifically motivates our research, concerns mixing regular items and sponsored ads. Whenever a user enters a search phrase, they are presented with a "sponsored listing", i.e., a listing that consists of a mixture of non-sponsored (organic) offers and ads. A major challenge is to interleave ads with organic search results in a way that balances user experience and the platform's ad revenue. To determine the positions of ads on a listing, we adapt a method inspired by~Ge~et~al.~\cite{ge2020relevance}. This method requires ranking scores with probabilistic interpretation which could be served by a calibrated ranker.

\subsection{Paper Contribution}

The contributions of this paper are as follows:
\begin{enumerate}
\item we show that it is possible to convert raw ranker scores to well-calibrated click probabilities using a flexible, plug-and-play post-hoc calibration technique: Multi-Layer Platt (MLPlatt)
\item we demonstrate how MLPlatt outperforms other state-of-the-art calibration methods in crucial metrics by leveraging context awareness 
\end{enumerate}

\section{Related work}

Learning to Rank (LTR)~\cite{liu2009learning} is the application of machine learning which focuses on creating models capable of predicting the most relevant order of a set of items given the query and other potentially relevant contextual information, such as user preferences. Over the years, three primary approaches have emerged to tackle this problem \cite{Cao2007LearningTR,li2011short}: treating each item individually and solving a regression or classification problem (the point-wise approach), focusing on ordering pairs of items (pair-wise), or considering the entire list of items for a given query (list-wise). Each approach requires differently structured data. This problem, in combination with non-differentiable evaluation ranking metrics like Normalised Discounted Cumulative Gain (NDCG), led to the development of sophisticated loss functions which aim to optimise ranking metrics indirectly. These loss functions include RankNet, LambdaRank, LambdaMART~\cite{burges2010ranknet} or the LambdaLoss Framework \cite{Wang2018TheLF}.  However, ranker model outputs trained with such loss functions are intrinsically uncalibrated and cannot be interpreted as probabilities.

Calibration methodologies aim to align predicted probabilities with true posterior probabilities, enabling a more accurate representation of uncertainty in predictions. The significance of calibration has been recognised across various domains, including image classification and CTR prediction.
For an excellent review of existing techniques, the reader is referred to the work by Pan~et~al.~\cite{Pan2019FieldawareCA}.

\begin{sloppypar}
Calibration methods have previously been applied in ranking for various purposes. To support the continuous training of ranking models, a calibrated objective function incorporating a reference element has been proposed in \cite{Yan2022ScaleCO}. A calibrated ranking multi-objective loss function was proposed in \cite{Bai2022RegressionCL}. Although their loss also includes a list-wise loss term, utilising a sigmoid transformation to make it better calibrated, it still puts more weight on the point-wise part to keep it well calibrated with marginal NDCG gains  \cite{Bai2022RegressionCL}. To the best of our knowledge, no prior research has explicitly focused on calibrating ranker models to make the outputs interpretable as CTR probabilities.
\end{sloppypar}

\begin{sloppypar}
To evaluate the MLPlatt framework proposed in this study, we selected some post hoc non-parametric (Isotonic Regression \cite{Jiang2011SmoothIR}), parametric (Platt \cite{Platt1999ProbabilisticOF}), and hybrid (DESC \cite{Yang2024DeepES} and ConfCalib \cite{Zhao2024ConfidenceAwareMM}) calibration methods. Furthermore, we benchmark against a ranker trained with Regression-Compatible Ranking loss (RCR \cite{Bai2022RegressionCL}). Such rankers, in contrast to those trained with LambdaLoss, output calibrated scores and are supposed to achieve a good balance between regression and ranking metrics.
\end{sloppypar}

\section{Methods}

\subsection{MLPlatt}\label{sec:calibration}

We will use the following notation for the input features: an item will be represented by $x_{item}$ and the ranking context (e.g. query, user country or user shopping history) by $x_{ctx}$.

Each item $x_{item}$ is represented by its multidimensional feature vector. This feature vector may, after applying a suitable embedding procedure, encompass both content-based and behaviour-based attributes. Content-based attributes describe the intrinsic features of the item, such as product name, price, or brand. Behaviour-based features describe how users interact with an item and can include metrics such as the number of purchases or the number of clicks over a time window.

The ranking context $x_{ctx}$ is represented by query-level features, such as the search phrase (e.g. "running shoes"), as well as user features,
for example, age or purchase history. It also includes categorical fields, which are important from a decision-making perspective. For example, for our proprietary dataset (see Section~\ref{sec:datasets} for details)
we focus on the device used to access the service (such as a smartphone app or a web browser).
For the AliExpress dataset, we focus on the country feature, following the methodology in \cite{Zhao2024ConfidenceAwareMM}. 

This section introduces the proposed solution called MLPlatt, a Multi-Layer Platt architecture
which can be used to convert ranker model scores into interpretable CTR probabilities.

\subsubsection{Model architecture}

\begin{figure}[t]
\centering
\includegraphics[width=\linewidth]{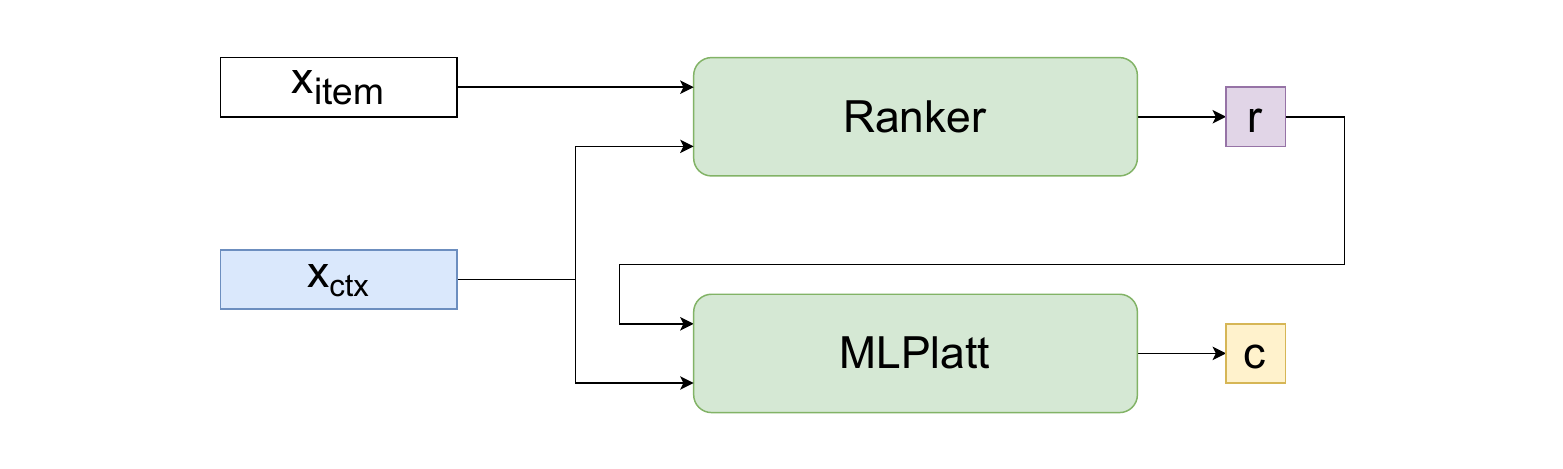}
\caption{High-level overview of the calibrated ranker model. Item ($x_{item}$) and context ($x_{ctx}$) features are passed through the ranking model to produce an uncalibrated score $r$. This score is then passed together with the context features through the MLPlatt model to output a calibrated score $c$.}
\label{fig:calibration-model-overview}
\end{figure}

\begin{figure}[t]
\includegraphics[width=\linewidth]{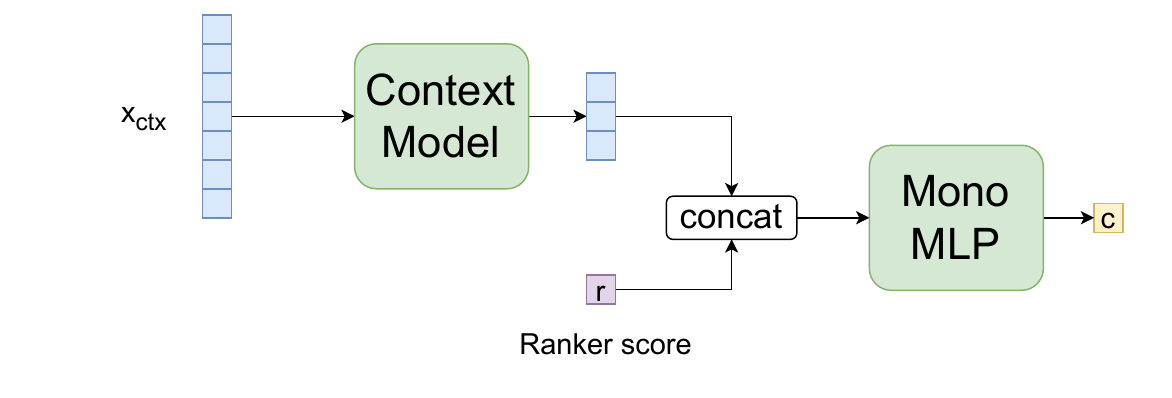}
\caption{MLPlatt model structure. The context vector $x_{ctx}$ is processed through the context model to produce an embedding. The embedding is concatenated with the ranker score, which is then passed to the MonoMLP model.}
\label{fig:mlplatt-model}
\end{figure}

An overview of the calibrated model structure is shown schematically in Figure~\ref{fig:calibration-model-overview}. The ranker model $f$ is a function of $x_{ctx}$ and $x_{item}$ and outputs item relevance $r$, i.e. $r = f(x_{ctx}, x_{item})$. We define the ranking model calibration problem as finding suitable weights of the calibration model $g$, which transforms the ranker score and context features into a calibrated probability of click, denoted by $c = g\left(r, x_{ctx}\right)$. We require the calibration model to preserve item ordering output by the ranker, hence we need the function $g$ to be monotonically increasing with respect to the ranker score $r$.

Note that for the calibration to maintain the ordering of items imposed by the ranker, item features $x_{item}$ are not used as inputs to the calibration model. Unlike in typical calibration problems, however, in ranking calibration we can take full advantage of the listing-level context features $x_{ctx}$. Listing-level features do not affect item ordering within a single listing, but may help the CTR predictions anchor at a correct value. Intuitively, the $x_{ctx}$ input to the calibration model allows it to learn differences in average CTRs between different contexts (users, phrases, devices, etc.), while the ranker score $r$ allows it to differentiate between items within the same listing.

The structure of the MLPlatt model is outlined in Figure~\ref{fig:mlplatt-model}.
Initially, the context features $x_{ctx}$ are passed to the Context Model to produce
the context embedding. The context embedding is then concatenated with the uncalibrated
ranker score $r$. The concatenated vector is then passed through
the MonoMLP model --- an MLP model constrained to be monotonic with respect
to the ranker score. This is achieved by adding the monotonicity constraint to the loss
function during training, as detailed in Section~\ref{sec:training}.
The final layer of the MonoMLP contains a sigmoid activation function
that outputs the final calibrated score $c$.

E-commerce ranking models serve hundreds of queries per second and work under strict
latency constraints. While the calibration model could in principle be arbitrarily complex, we
deliberately aim for MLPlatt to be a lightweight model. In practice, this means that
rather than passing the untransformed $x_{ctx}$ vector
to the context model, we pass embeddings computed by the ranker. In the case of the Allegro data, this
means passing, e.g. an embedded search phrase extracted from a hidden layer of the ranker.


\subsubsection{Training procedure}
\label{sec:training}

To train a calibrated ranker, we employ a two-step training procedure. First, a backbone ranker model is trained on the dataset using a ranking-specific loss function. Then, a calibration model is trained on the same dataset using a point-wise loss function. In the second stage, the calibration model uses ranker scores and contextual features as inputs, but discards the item-specific features.

The ranker model is trained on the training set for several epochs. The $N$ ranker scores, $r^{i}$, for each query-item pair are then recorded and, together with the query-level context features, $x_{ctx}^i$, and target binary click labels, $c_{true}^i$, form the training set $\left(\left(r^{i}, x_{ctx}^i\right), c_{true}^i\right)_{i=1 \dots N}$ for the calibrator.

To enforce the monotonicity of the calibrator, we employ a method similar to the ones proposed in~\cite{gupta2019incorporatemonotonicitydeepnetworks,monteiro2022monotonicityregularizationimprovedpenalties}.
Define $d^i = \tfrac{\partial c}{\partial r}\big(r^{i}, x_{ctx}^i\big)$ to be the partial derivative of the calibrated score with respect to the ranker score. We then augment the BCE loss with a monotonicity-enforcing penalty loss over the empirical train dataset distribution:
\[
    L_{Mono} (d) =
    \frac{1}{N} \sum_{i=1}^{N} \max \left(0, - d^i \right)
\]
This component of the loss function effectively penalises the negative derivative of the calibrator with respect to the ranker score, thus enforcing the calibration model to be a monotonically increasing function of the ranker score. This encourages preservation of the ranking order by the calibration model.

The final loss used for MLPlatt model training then becomes
\begin{equation}\label{eq:loss-function}
    L_{Calib}\left(c_{true}, c, d; \theta \right) =
    L_{BCE}(c_{true}, c) + \theta L_{Mono} (d), 
\end{equation}
where $L_{BCE}$ is the BCE loss and $\theta$ is the weight used to penalize non-monotonicity. We perform an ablation study with respect to the monotonicity penalty $\theta$ in Section~\ref{sec:theta-sensitivity}.

\subsubsection{Integrating MLPlatt with existing pipelines}

We note that the MLPlatt model can be easily incorporated into existing
ranking pipelines. MLPlatt is agnostic with respect to the underlying ranking
model architecture, and it reuses the contextual features which should be readily
available. The only real requirement is that the ranking scores from the underlying
ranking model are computed for the entire dataset and that a sequential two-step training
procedure must be used to first train the ranker, followed by the calibration model.
This requirement is shared among all post hoc calibration methods and is not specific to MLPlatt.

\section{Experiments}

General dataset information and model architecture details are
outlined in Table~\ref{tab:model-details}. The layer sizes were fine-tuned by a manual grid search over a small range of values.

\begin{table}[t!]
\caption{Details of the experiment setup: dataset features and MLPlatt
model hyperparameters. Dash-separated numbers in Context Model and MonoMLP.
Columns denote layer sizes in MLP networks. "Identity" denotes passthrough
layer which returns the input unchanged.}
\centering
\begin{tabular}{l|rr|rr}
Dataset & Listings & $x_{ctx}$ dim. & Context Model & MonoMLP \\
\toprule
Allegro & 200M & 69 & 32-16-8 & 8-8-8-1 \\
AliExpress & 5M & 8 & Identity & 8-8-8-1 \\
\end{tabular}
\label{tab:model-details}
\end{table}

\subsection{Datasets}
\label{sec:datasets}

\subsubsection{Allegro} This proprietary dataset consists of 200M search listings sampled from a single year from an e-commerce platform \url{http://www.allegro.pl}.
The listings represent typical traffic and cover all four devices that can be used to
access the platform. We are mainly interested in achieving good calibration for each device separately, hence we treat it as a calibration \emph{field} (using the terminology of Yang~et~al. \cite{Yang2024DeepES}).
The data consists of various user, item, and contextual features, as well as columns indicating whether an item was clicked by the user, which we use as the target variable.

The ranking model for the Allegro dataset is the DeText \cite{guo2020detext} trained using LambdaLoss \cite{Wang2018TheLF}. We note that training with LambdaLoss results in a ranker which is uncalibrated by construction. For MLPlatt, we used
an MLP with layers of size 32-16-8 for the Context Model and MonoMLP with layers of size 8-8-8-1.

\subsubsection{AliExpress}
\begin{sloppypar}
The AliExpress dataset was gathered from real-world traffic logs of the search system in AliExpress\footnote{Available at https://tianchi.aliyun.com/dataset/74690} \cite{Ma2018EntireSM}. Similarly to \cite{Zhao2024ConfidenceAwareMM} we use data from 4 countries, excluding Russia. To get comparable results to the Allegro dataset, we use clicks as the target variable. We split the dataset into
train and test subsets using a 2:1 ratio, as suggested by the dataset documentation. The training dataset then consists
of roughly 4.6M listings. We further filter the dataset to listings suitable for training a ranking model, i.e.
with at least one clicked item. 
Both the ranker and post-hoc calibration methods are trained on the training subset and evaluated on the test subset. As the target fields, we choose the \texttt{country} feature.
\end{sloppypar}

The AliExpress dataset does not contain search phrase, so the ranker model used was a fully-connected network with three layers, trained with the LambdaLoss \cite{lambdaloss} objective function. For MLPlatt we used MonoMLP with layers of size 8-8-8-1. No Context Model was used for simplicity, as the context feature $x_{ctx}$ has only 8 dimensions.

\subsection{Metrics}
\label{sec:metrics}

In this study, we aim to optimise jointly for the quality of calibration and the quality of ranking.
We are interested in achieving high-quality calibration not only globally, but also within
strata corresponding to different values of the target-sensitive fields --
device for the proprietary dataset and country for the AliExpress dataset. We follow the methodology of Pan~et~al. \cite{Pan2019FieldawareCA} and use field-aware calibrating metrics.

We adapt the notation conventions from \cite{Pan2019FieldawareCA}: let $\mathcal{Z}$ denote possible
values of a discrete field (e.g. device type). The entire dataset $\mathcal{D}$ can then be partitioned into disjoint sets
$\mathcal{D}_z$ of observations having $z\in \mathcal{Z}$ as the value of the field in question.
Moreover, $c_{true}$ denotes the true binary labels, while $c$ denotes the calibrated model predictions.

Our main calibration metric is the field-level expected calibration error at $M$ bins (F-ECE@M). It is defined as the average
of ECE(z)@M calculated separately for each field value $z$, weighted by the number of observations in that field.
For each subset $\mathcal{D}_z$ the ECE(z)@M is computed by binning the observations into M same-sized bins
$B^{z}_{m}$ using the quantiles of the predicted scores and computing the mean difference between the average
model prediction and a fraction of positive observations in each bin:
\[ \text{ECE(z)@M} = \frac{1}{M} \sum_{m=1}^M \frac{\left| \sum_{B^z_m} \left(c_{true} - c\right) \right|}{|B^z_m|}  \]
Intuitively, ECE(z)@M measures how far the model's calibration curve is from the perfect calibration. 
The F-ECE@M can then be computed as:
\[
    \text{F-ECE@M} = \frac{1}{|\mathcal{D}|} \sum_{z \in \mathcal{Z}} |\mathcal{D}_z|~\text{ECE(z)@M}
\]
In this paper, we use $M=20$ for the number of bins and will henceforth drop the $@M$ suffix.

In addition to F-ECE, we report the Log-Loss and AUC, two typical measures reported for CTR models.
We use the Normalised Discounted Cumulative Gain (NDCG) \cite{jarvelin2002cumulated} as a ranking quality measure, a standard choice for evaluating information retrieval systems.

\subsection{Comparative Methods}

To provide a wide range of comparisons, we tested the following model calibration approaches: Platt and Isotonic regression (commonly used parametric and non-parametric baselines), as well as ConfCalib and DESC (recent state-of-the-art models designed explicitly for multi-field calibration of CTR models). We briefly describe each in this section, while the details can be found in the references provided.

\bmhead{Platt Scaling}
Platt Scaling \cite{Platt1999ProbabilisticOF} is a basic solution to the calibration problem, learning a linear transformation of the output followed by a sigmoid function. The parameters of the transformations are estimated using maximum likelihood.

\bmhead{Isotonic Regression}
The solution based on Isotonic Regression is similar to \cite{Deng2020CalibratingUR}, where authors begin with binning (for the calibration curve), then train the Isotonic model using the binned representation, and use linear interpolations for prediction.

\bmhead{ConfCalib}

\begin{sloppypar}
Confidence-Aware Multi-Field Model Calibration (ConfCalib) \cite{confcalib} is a non-parametric calibration method that takes into account the confidence of field data observations. The method considers user feedback as a binomial distribution of field values and uses Wilson confidence intervals to compute deviations of model predictions. These deviations are made smaller via a non-linear transformation, and subsequently used for computing new confidence intervals, which are then used to calibrate the predictions via scaling. Since the authors did not provide a source code, we re-implemented the solution based on the paper.
\end{sloppypar}

\bmhead{DESC}

Deep Ensemble Shape Calibration (DESC\footnote{https://github.com/HaoYang0123/DESC}) \cite{Yang2024DeepES} dissects the calibration problem into two distinct sub-problems: shape calibration and value calibration, creating two parallel calibrators. 
The Shape Calibrator is responsible for assigning appropriate basis calibration functions to mitigate the issues of over- and under-estimation across all output intervals. Conversely, the Value Calibrator concentrates on correcting over- and under-estimation for individual samples.

\section{Results}

\subsection{Calibration experiments}

The outcomes of the various calibration techniques on the  model trained on the proprietary dataset are shown in Table~\ref{tab:allegro-benchmark}. It can be seen that MLPlatt achieves the best F-ECE, log loss, and AUC scores without adversely impacting NDCG.

Analogous results for the AliExpress public dataset are depicted  in Table~\ref{tab:aliexpress-benchmark}. Similarly as in the proprietary dataset, MLPlatt leads in the F-ECE and AUC metrics. DESC achieves similar log loss (change is at the 5th decimal place, hence not shown), but at the cost of a significant decrease in the NDCG metric.

Across both datasets, MLPlatt is the only model capable of lifting the regression and calibration metrics without hurting the ranking. We believe that its main advantage over other calibration techniques is its effective usage of the listing-level contextual features.

\begin{table}[t!]
\caption{
    Calibration and ranking metrics for different calibration methods on the Allegro dataset. Best values are marked with \textbf{boldface}. Statistically significant difference (compared to MLPlatt) with p-value smaller than $0.01$ is denoted with an asterisk.
}\label{tab:allegro-benchmark}
\centering
\begin{tabular}{l|rrrrrr}
    \toprule
    Calibration method & F-ECE & LogLoss & NDCG & AUC \\
    \midrule
    Platt~\cite{Platt1999ProbabilisticOF}            &       $^*$0.0277 &       $^*$0.2970 &  \textbf{0.5082} &       $^*$0.5671 \\
    Smoothed Isotonic Regression~\cite{Deng2020CalibratingUR} &       $^*$0.0263 &       $^*$0.2965 &           0.5081 &       $^*$0.5671 \\
    ConfCalib~\cite{confcalib}                       &       $^*$0.0132 &       $^*$0.2943 &  \textbf{0.5082} &       $^*$0.6142 \\
    DESC~\cite{Yang2024DeepES}                       &       $^*$0.0025 &       $^*$0.2916 &       $^*$0.5067 &       $^*$0.6225 \\
    MLPlatt ($\theta=1$)                             &  \textbf{0.0021} &  \textbf{0.2896} &  \textbf{0.5082} &  \textbf{0.6350} \\
    \bottomrule
\end{tabular}
\end{table}

\begin{table}[t!]
\caption{
    Calibration and ranking metrics for different calibration methods on the AliExpress dataset. Best values are marked with \textbf{boldface}. Statistically significant difference (compared to MLPlatt) with p-value smaller than $0.01$ is denoted with an asterisk.
}\label{tab:aliexpress-benchmark}
\begin{tabular}{l|rrrrrr}
\toprule
Calibration method & F-ECE & LogLoss & NDCG & AUC \\
\midrule
Platt~\cite{Platt1999ProbabilisticOF}            &       $^*$0.0056 &       $^*$0.3044 &  \textbf{0.5481} &       $^*$0.6469 \\
Smoothed Isotonic Regression~\cite{Deng2020CalibratingUR} &       $^*$0.0709 &       $^*$0.3698 &  \textbf{0.5481} &       $^*$0.6469 \\
ConfCalib~\cite{confcalib}                       &       $^*$0.0752 &       $^*$0.3865 &  \textbf{0.5481} &       $^*$0.5880 \\
DESC~\cite{Yang2024DeepES}                       &       $^*$0.0032 &       $^*$0.3043 &       $^*$0.5469 &       $^*$0.6477 \\
MLPlatt ($\theta=1$)                             &  \textbf{0.0028} &  \textbf{0.3043} &  \textbf{0.5481} &  \textbf{0.6480} \\
\bottomrule
\end{tabular}
\centering
\end{table}

\subsection{Ablation study}

We perform an ablation study on the Allegro dataset for two reasons. Firstly, it consists
of 200M listings and is much larger in comparison to the AliExpress 4.6M dataset.
Furthermore, our dataset contains the search phrase which we use as the contextual feature,
which we consider typical for ranking problems arising in e-commerce.

Due to the modular nature of MLPlatt as seen in Figure~\ref{fig:mlplatt-model}, it is
possible to examine the impact of each individual module, such as the context model and
MonoMLP, separately. We examine the impact of each component on both the ranking and CTR
metrics. The results are shown in Table~\ref{tab:allegro-ablation}.

We examine the MLPlatt model but with the Context Model replaced by an identity layer ($x \mapsto x$). In this case, we see that the F-ECE metric is worse than for a full MLPlatt model. The $\theta=1$ penalty in the loss function is also not enough to enforce monotonicity and NDCG decreases. We conclude that the Context Model is crucial for MLPlatt to utilise the contextual features effectively.

Next, we consider MLPlatt with MonoMLP replaced by an MLP with a single layer. We see that this model is NDCG-neutral, but yields worse F-ECE than the MLPlatt with no Context Model.

We finally we benchmark against the original Platt \cite{Platt1999ProbabilisticOF} method. This is equivalent to not providing the context features $x_{ctx}$ and having a single-layer MonoMLP model in MLPlatt. As we can see all calibration and regression metrics are much worse than in the case of full MLPlatt.

We see that the full MLPlatt model is best both in terms of calibration and ranking metrics. 

\subsection{Sensitivity to Monotonicity Penalty}\label{sec:theta-sensitivity}

In this section we discuss the sensitivity of the results to the weight $\theta$ in Equation~\eqref{eq:loss-function}. We sampled 100,000 listings from Allegro validation dataset and calculated spearman rank correlation for each listing between the uncalibrated scores and scores calibrated using MLPlatt for different values of the monotonicity penalty $\theta$. We then counted the fraction of listings for which the correlation coefficient was smaller than 0.99. As seen in Table~\ref{tab:theta-sensitivity}, switching off the penalty results in changes in ordering in 2.20\% of listings. That fraction decreases as $\theta$ increases and we note that setting $\theta = 1$ is enough to ensure that MLPlatt is a monotonic transformation.

\begin{table}[t!]

\caption{Fraction of listings whose order is affected by MLPLatt calibration model depending on the monotonicity penalty $\theta$ in the loss function.}\label{tab:theta-sensitivity}
\centering
\begin{tabular}{r|r}
\toprule
$\theta$ & \% of misordered listings \\
\midrule
$0$ & $2.20\%$ \\
$1\mathrm{e}{-4}$ & $1.58\%$ \\
$1\mathrm{e}{-3}$ & $1.52\%$ \\
$1\mathrm{e}{-2}$ & $0.46\%$ \\
$1$ & $0.00\%$ \\
\bottomrule
\end{tabular}

\end{table}

\begin{table}[t!]
\caption{
    Influence of components of MLPlatt on the calibration and ranking metrics. Best results are marked with \textbf{boldface}.
}\label{tab:allegro-ablation}
\begin{tabular}{l|rrrrrr}
    \toprule
    Calibration method & F-ECE & LogLoss & NDCG & AUC \\
    \midrule
    Platt~\cite{Platt1999ProbabilisticOF}            &           0.0277 &           0.2970 &  \textbf{0.5082} &           0.5671 \\
    MLPlatt (No Context Model)                       &           0.0082 &           0.2921 &           0.4942 &           0.6250 \\
    MLPlatt (No MonoMLP)                             &           0.0061 &           0.2904 &  \textbf{0.5082} &           0.6328 \\
    MLPlatt                                          &  \textbf{0.0021} &  \textbf{0.2896} &  \textbf{0.5082} &  \textbf{0.6350} \\
    \bottomrule
\end{tabular}
\centering
\end{table}

\subsection{Comparison with Regression-Compatible Ranking Loss}

In this section, we compare the results on the Allegro dataset with the Regression-Compatible Ranking loss (RCR)~\cite{Bai2022RegressionCL}. RCR uses a weighted average of a point-wise loss function and a special list-wise loss function. By adjusting the weight of each term, one can balance the trade-off between CTR prediction and ranking objectives. 

Training ranker with RCR is a convenient alternative to post hoc calibration methods as it does not require two-stage training. We found, however, that even though the NDCG is comparable to LambdaLoss-trained rankers, the F-ECE score is much worse for RCR than for a ranker trained using LambdaLoss and then calibrated with MLPlatt. The results are shown in Table~\ref{tab:allegro-rcr}. Interestingly, MLPlatt yields lower AUC than some of the RCR models which suggests that RCR-trained rankers are better at classification but suffer from poor calibration.

\begin{table}[t!]
\caption{
    Comparison of LambdaLoss+MLPlatt and Regression-Compatible Ranking loss (RCR) \cite{Bai2022RegressionCL} calibrated rankers. Higher $\alpha$ means more focus on ranking objective in the RCR loss. Best values are marked with \textbf{boldface}. Statistically significant differences with respect to MLPlatt with p-value smaller than $0.01$ are marked with asterisk.
}\label{tab:allegro-rcr}
\centering
\begin{tabular}{l|rrrrrr}
    \toprule
    Calibration method & F-ECE & LogLoss & NDCG & AUC \\
    \midrule
    RCR ($\alpha=1\mathrm{e}{-3}$)                   &       $^*$0.0376 &       $^*$0.3060 &       $^*$0.5058 & $^*$\textbf{0.6501} \\
    RCR ($\alpha=1\mathrm{e}{-2}$)                   &       $^*$0.0394 &       $^*$0.3093 &       $^*$0.5061 &       $^*$0.6477 \\
    RCR ($\alpha=1\mathrm{e}{-1}$)                   &       $^*$0.0380 &       $^*$0.3124 &           0.5080 &           0.6363 \\
    MLPlatt                                          &  \textbf{0.0021} &  \textbf{0.2896} &  \textbf{0.5082} &           0.6350 \\
    \bottomrule
\end{tabular}

\end{table}

\section{Discussion}

The results we obtained for the proprietary Allegro and public AliExpress datasets are in line with other results regarding field-aware calibration methods presented in previous works, e.g. \cite{Pan2019FieldawareCA,Yang2024DeepES,confcalib}. Classic calibration
methods, such as Platt scaling or Isotonic Regression, are incapable of producing reliable
calibration consistently across sensitive fields (e.g. device type or country). Notably, our proposed method also outperforms two new, state-of-the-art approaches, namely DESC \cite{Yang2024DeepES} and ConfCalib \cite{Zhao2024ConfidenceAwareMM}. Our proposed MLPlatt calibration model achieves the best calibration metric F-ECE among all tested methods, Log-Loss and AUC. 

Furthermore, MLPlatt can be constrained to be a monotonically increasing function of the ranker score by using a modified loss function $L_{calib}$. In practice, this allows replacing the ranker with a more interpretable, ranking-compatible CTR model without negatively affecting ranking metrics.

We note that, apart from MLPlatt, none of the other methods tested were designed specifically for post hoc ranking model calibration. They do not, therefore, fully leverage the contextual information ($x_{ctx}$) which can be used when calibrating rankers in the way MLPlatt does. 

We aimed for our methods to be NDCG-neutral and this was successfully achieved by MLPlatt.
Using a MonoMLP module and a modified loss function, we can enforce the monotonicity of the CTR prediction with respect to the ranker score, while at the same time achieving far greater expressivity by using a more complex neural network as the calibration head.


Moreover, we note that MLPlatt was purposefully designed to be lightweight when compared to the backbone ranking model. This means that using it in place of the original ranker produces a negligible latency overhead. Training MLPlatt is essentially training a small classification model; therefore, our approach scales well, even with commercially sized datasets.

\section{Conclusions}

In this paper, we propose MLPlatt, a novel framework to transform uncalibrated ranker predictions into CTR probabilities while preserving the ranking order. The method was demonstrated to be superior to several other strong baseline approaches on two distinct e-commerce datasets. Specifically, it successfully addressed the challenge of field-level calibration, showing that including context features significantly enhances calibration performance. Future work could explore opportunities to use calibrated ranking models in place of CTR models.

\bibliography{mlplatt}

\end{document}